\documentclass[3p,twocolumn]{elsarticle}
\usepackage{amssymb}
\usepackage{amsmath}
\usepackage{lineno}
\usepackage[T1]{fontenc}
\usepackage[utf8]{inputenc}
\usepackage[english]{babel}
\journal{Chaos Solitons and Fractals}

\begin{document}

\begin{frontmatter}

\title{Effects of a pestilent species on the stability of cyclically dominant species}
\author[p]{D. Bazeia}
\author[p]{M. Bongestab}
\author[m]{B.F. de Oliveira}
\author[e]{A. Szolnoki}
\address[p]{Departamento de Física, Universidade Federal da Paraíba, 58051-970 João Pessoa, PB, Brazil}
\address[m]{Departamento de Física, Universidade Estadual de Maringá, 87020-900 Maringá, Paraná, Brazil}
\address[e]{Institute of Technical Physics and Materials Science, Centre for Energy Research, P.O. Box 49, Budapest H-1525, Hungary}
\begin{abstract}
Cyclic dominance is frequently believed to be a mechanism that maintains diversity of competing species. But this delicate balance could also be fragile if some of the members is weakened because an extinction of a species will involve the annihilation of its predator hence leaving only a single species alive. To check this expectation we here introduce a fourth species which chases exclusively a single member of the basic model composed by three cyclically dominant species. Interestingly, the coexistence is not necessarily broken and we have detected three consecutive phase transitions as we vary only the invasion strength of the fourth pestilent species. The resulting phases are analyzed by different techniques including the study of the Hamming distance density profiles. Some of our observations strengthen previous findings about cyclically dominant system, but they also offer new revelations and counter-intuitive phenomenon, like supporting pestilent species may result in its extinction, hence enriching our understanding about these very simple but still surprisingly complex systems.
\end{abstract}
\begin{keyword}
\texttt cyclic dominance \sep biodiversity \sep pestilent predator
\end{keyword}

\end{frontmatter}

\section{Introduction}

Biodiversity is a subject of complex nature, but it still offers theoretical challenges to understand it in the framework of evolutionary game theory \cite{1996-Sinervo-Nature-380-240,2002-Kerr-N-418-171,2004-Kirkup-Nature-428-412,2007-Reichenbach-N-488-1046,DB3,DB2,DB1}. Rock-paper-scissors game-like cyclic dominance seems to be an adequate explanation why not just a singular victor of an evolutionary process survives, and this mechanism inspired a huge research activity in the last decades \cite{2018-Dobramysl-JPA-51-063001,2020-Szolnoki-EPL-131-68001,2007-Szabo-PR-446-97,2014-Szolnoki-JRSI-11-0735}.

The lack of transitivity in the rank of competing species produces several surprising effects \cite{mobilia_g16,nagatani_jtb19,szolnoki_pre16,brown_pre19,park_c18c,garde_rsob20,palombi_epjb20,nagatani_c20,szolnoki_srep16b}. For example, if we modify the invasion rates in the circle and weaken the efficiency of a predation process then the one who benefits the most from this intervention will be the affected, hence seemingly weakened predator. This is the so-called survival of the weakest phenomenon \cite{2001-Frean-PRSLB-268-1323,tainaka_pla93} which has been studied thoroughly in several different systems during the years and confirmed its broad robustness \cite{Berr-PRL-102-048102,blahota_epl20,2019-Avelino-PRE-100-042209,szolnoki_csf20b,nagatani_pa19b,2020-Avelino-PRE-101-062312,szolnoki_pre17,park_c18,2020-Bazeia-CSF-141-110356,wang_z_pre14b}. Notably, in a recent microbial experiment Liao~{\it et~al.} implemented an experimental investigation, unveiling the survival of the weakest in the non-transitive asymmetric interactions among strains of {\it Escherichia coli} populations of bacteria \cite{Liao2020}. This work studied a microbial community with three strains of bacteria that cyclically interact through the inhibition of protein production, the digestion of genomic deoxyribonucleic acid, and the disruption of the cell membrane. The results found intrinsic differences in these three major mechanisms, ending in an unbalanced community that was dominated by the weakest strain.

An interesting link between chaotic and cyclically dominant system was established by a recent work in which the authors proposed a novel procedure to identify the presence of chaos in stochastic simulations that are used to study cyclic models of the rock-paper-scissors game \cite{2017-Bazeia-SR-7-44900}. This investigation used the Hamming distance concept \cite{Hamming} to measure the distance between two states that are slightly different initially, but evolve under the very same stochastic rules that controls the time  evolution of the model. The behavior remind us very much of the butterfly effect, related to the sensitive dependence on initial conditions, revealing that a small change in one state can result in large differences in the time evolution of the system. The Hamming distance density (HDD) was further studied in generalized rock-paper-scissors models with several species and showed the very same qualitative evolution, hence suggesting an universal behavior \cite{2017-Bazeia-EPL-119-58003}.

Summing up, based on our previous observations, there is a consensus that to maintain biodiversity needs the presence of all competitors, because the extinction of a prey can break this delicate balance which eventually leads to the dying out of the related predator species, leaving only a single species alive. 

Therefore we may expect that an external predator can jeopardize the stability of the whole system. This idea motivated a model study where three distinct species evolve under the action of another species, a so-called apex predator who supervises all the three other species, without being predated by any of them \cite{2017-Souza-Filho-PRE-95-062411}. The main result showed that the apex predator spreads uniformly in space, contributing to destroy the spiral patterns, diminish the average size of the competing clusters, but keeping biodiversity alive. The model was further studied in the case where the apex predator decaying ratio is varied to unveil different phases of the model \cite{2018-Bazeia-EPL-124-68001}.

In the above mentioned cases, however, the fundamental symmetry of the basic model was not broken, because the external predator attacks all competing species in the circle. But what happens if only one species is in danger? Does the sensitivity of a member make the biodiversity fragile? Motivated by these questions and the above mentioned works we here study the standard May-Leonard model of three species where beside the usual reproduction, competition and mobility processes we introduce another species to predate only one of the cyclical species. These new individuals characterize a pestilent population which threaten indirectly the whole system. Notably, our present model is conceptually different from previous studies, where the authors focused on systems with three, four and more species that compete cyclically; see, e.g., Refs. \cite{2018-Dobramysl-JPA-51-063001,2011-Durney-PRE-83-051108,2013-Knebel-PRL-110-168106}. In these works, it was shown that although the survival of the weakest appears in a three-species system which evolves cyclically (in the $3SS$ model), it may not show up in the case of four species, called $4SS$ model.

\section{Threatening biodiversity by a pestilent predator}

We here study a spatial model where four species are distributed on a $N \times N$ square lattice where periodic boundary conditions are applied. The first three species, $1,2,3$, who are marked by red, blue, and yellow colors respectively, evolve under the standard rules of May-Leonard model \cite{2017-Brown-PRE-96-012147}. Accordingly, the elementary processes are mobility with probability $m$, reproduction with probability $r$, and competition with probability $p$. Crucially, the latter is characterized by the non-hierarchical rules of the well-known rock-paper-scissors game, in which scissors cuts paper, paper wraps rock and rock breaks scissors in a cyclic manner. When a prey is eliminated, it leaves an empty site behind, which microscopic state is identified as $0$ and marked by white color. 

\begin{figure}[!htb]
	\centering
		\includegraphics[width= 3.5cm]{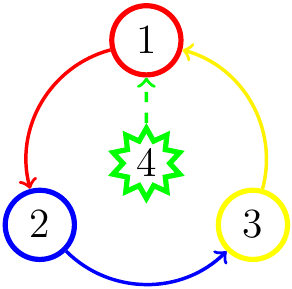}
	\caption{Schematic view of the model investigated in the present work. The solid red, blue and yellow arrows indicate competition process of the May-Leonard type model, while dashed green arrow shows a Lotka-Volterra model-type predation of the green species against the red species.}
	\label{fig1}
\end{figure}

The novelty of our model is the presence of a fourth species, who is represented by $4$ and denoted by green color. In stark contrast to previously used external predators, it only predates species $1$, the red one, and this predation process follows a Lotka-Volterra-type invasion rule \cite{2013-Knebel-PRL-110-168106}. In particular, a predation is followed by the reproduction process of the predator immediately. As we stressed, this external species is harmful to species 1 only and is neutral to the remaining two members of the cycle. Therefore it acts as a pestilent species, as a plague whose only role to kill the red population. A schematic view of the proposed model is illustrated in Fig.~\ref{fig1}. 

The numerical simulations starts with the preparation of an initial state, which follows by specifying the lattice size choosing $N$ as the linear size, and then distributing the species and empty sites in the lattice using standard random choices. This procedure establishes an initial state where red, blue, yellow, green and white sites are almost uniformly distributed in the square lattice. 
The time evolution follows as: we first access the lattice by randomly selecting a site, the active site, and one of its neighbors which is considered to be the passive site. Here we use the von Neumann neighborhood, in which each site $(i,j)$ has four neighbors, $(i\pm1,j\pm1)$, for $i$ and $j$ in the set $1,2,\dots, N$. If the active site is empty we start the selection step from the beginning. 
But if the chosen site is occupied by one of the members of the cycle, we then randomly select a microscopic process among mobility, reproduction, and competition with probabilities
$m, r$, and $p$ respectively. In agreement with previous model studies we here applied $r=p=(1-m)/2$ relation and implement it as follows: if mobility is chosen, the states of the active and passive sites change their positions. If reproduction is chosen and if the passive site is empty then it is occupied by the state (color) of the active site. Finally, if competition is chosen and the involved positions are occupied by different members of cyclical species then prey species dies out leaving an empty position behind. Moreover, if the active site is occupied by a green individual, we first necessarily move it, exchanging position with the passive neighbor already chosen.  After this motion, we select from two complementary processes which are chosen with probability $d$ and $q$, with $d+q=1$. If the first is chosen, the green individual is eliminated leaving the site empty. In case of the alternative choice we select another neighbor and if it is occupied by a red individual then it is replaced by a green one. Evidently, only just a red neighbor is eliminated, but a blue or a yellow neighbor remains intact. The elimination of red neighbor simply means that the predation and reproduction processes are executed in a single step, as it is assumed in Lotka-Volterra-type description. It is easy to see that the value of $q$ represents the invasion strength of the external species, while $d$ denotes the annihilation rate of this species. In a full simulation step we select each available lattice site once on average, hence the executions of the above described procedure $N^2$ times define a natural time unit, so-called Monte Carlo (MC) step of the simulation. Typically we employed $N=1000$ linear system size, by using $10^4$ independent initial states where the length of the simulation steps varied between $10^3$ and $10^6$ MC steps. To quantify the emerging phases we measure the average $\rho_i$ ($i = 0, \dots, 4$) density of all species in the stationary states. Here $i=1, \dots, 4$ refer to the related species defined by Fig.~1, while $i=0$ refers to the density of empty spaces.

Since only a single member is attacked externally, it would be a straightforward expectation that the stability of the whole biodiversity is in danger. Another key question to explore is how the weakening of red species will change the relations of the cyclic species.

\section{Results}

\begin{figure}[h!]
	\centering
		\includegraphics[width= 4.9cm]{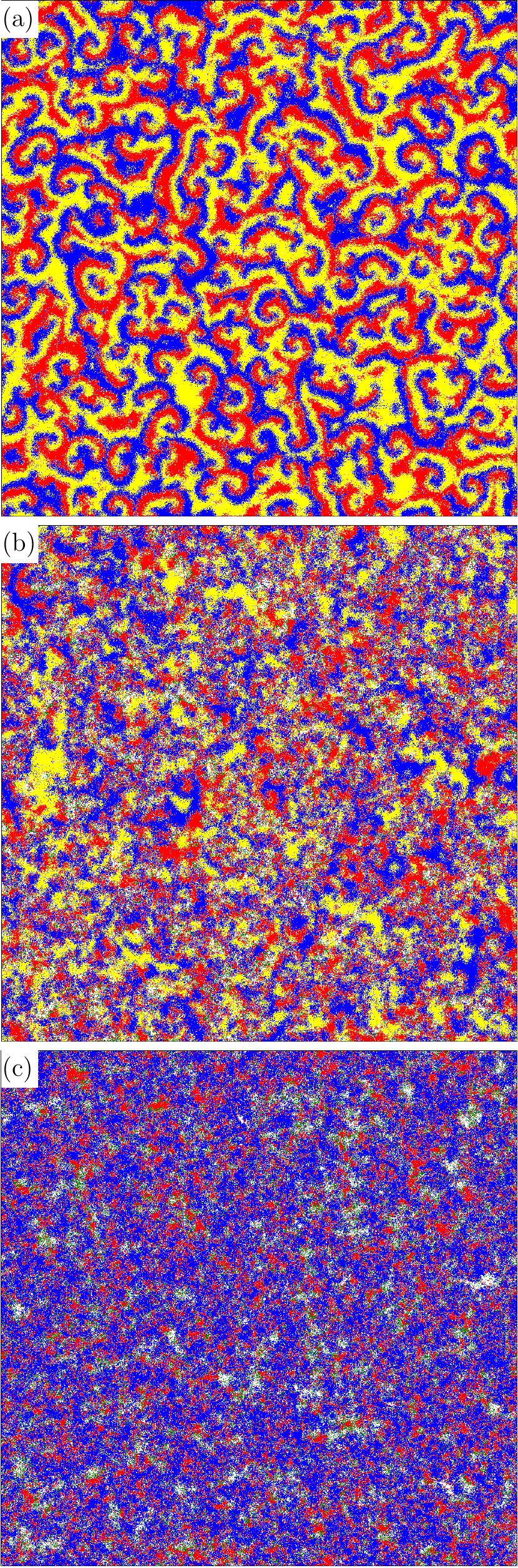}
	\caption{Representative snapshots of the time evolution of the model for three distinct values of $q$, after 10000 MC steps. Here we used the color codes for species introduced in Fig.~1, while white color shows empty lattice sites. In panels $\rm (a)$, $\rm (b)$, and $\rm (c)$ we depict the patterns obtained at $q=0.75$, $0.80$, and $0.85$, respectively. By only changing the $q$ invasion strength of pestilent species we can identify distinct phases, as discussed in detail in the main text.}
	\label{fig2}
\end{figure}

A representative overview about the system behavior is shown in Fig. \ref{fig2}, where we display three distinct snapshots obtained after 10000 MC steps. We use a linear lattice size $N=1000$, and for the three species red, blue and yellow we take the standard values of $m=0.50$ and $r=p=0.25$ throughout this work, but we vary the predation rate $q$ of the green species. In Figure~\ref{fig2} the three panels are displayed for three representative values of the predation rate $q$: $\rm(a)$ is for $q=0.75$, $\rm(b)$ for $q=0.80$, and $\rm(c)$ for $q=0.85$, which identify three distinct phases of the system. In panel~$\rm (a)$ green species cannot survive and we identify the standard behavior for the red, blue and yellow species: they evolve according to the basic cyclic dynamics by forming and maintaining typical spiral patterns \cite{2020-Szolnoki-EPL-131-68001,frey_pa10}. 
In panel~$\rm(b)$ the spiral patterns become less visible and we can detect a stable coexistence of all four species, with the blue species as the most abundant one. In the next phase, shown in panel~$\rm (c)$, the system is unable to maintain spiral patterns because one of its member, the yellow species, disappears. In this phase the dominance of blue species becomes even more visible.

\begin{figure}[t]
	\centering
		\includegraphics[width= 6.5cm]{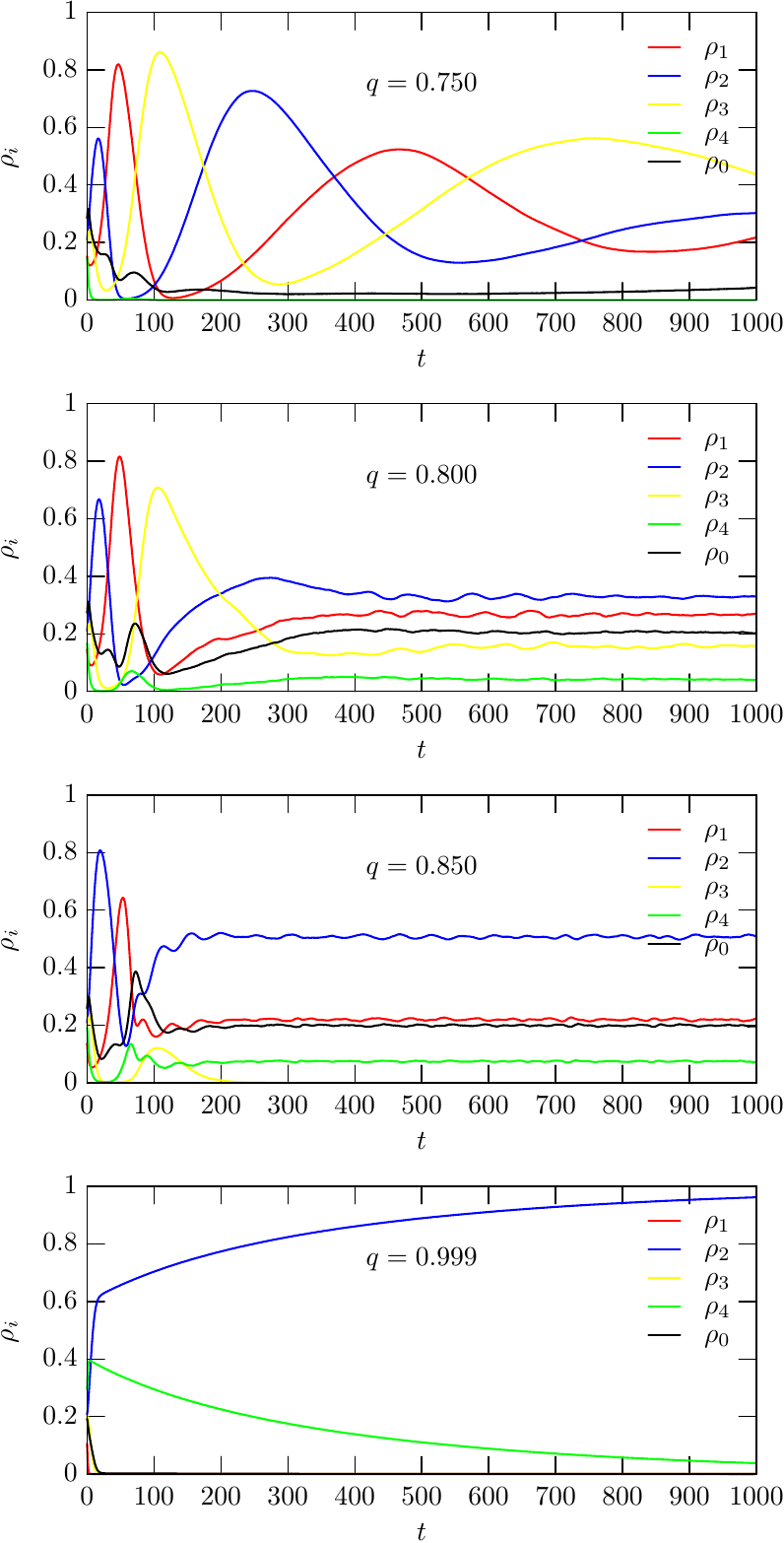}
	\caption{The abundance of the species and empty sites as a function of time up to 1000 MC steps when the evolution was launched from a random initial state. Different panels were obtained for different $q$ values as indicated in the legend. The last panel shows the evolution when the system evolves toward a full blue state.}
	\label{fig3}
\end{figure}

In order to better understand the governing process in the mentioned phases in Fig.~\ref{fig3} we present the related time evolutions of the densities of species and vacant sites when the evolution is launched from a random initial state. The first panel, obtained at $q=0.75$, suggests that the green pestilent species dies out very soon for low $q$ values and the system becomes equivalent to the traditional three-species cyclic model where after a strong initial transient the system gradually evolves to a state where all remaining species are present with equal densities (not shown in the panel). The reason why external invaders cannot survive can be explained as follows. As we mentioned, green species takes part in two microscopic processes. They invade neighboring red individuals with probability $q$, or die out spontaneously with probability $d = 1 - q$. An obvious threshold value would be $q_c = 0.5$ where these two processes are in a balance, but the neighboring site of a green individual is not necessarily occupied by a red one, hence the invasion, which would support the portion of green, cannot always be executed. It explains why we need a higher threshold value of $q$ or, in other words, even for $q>0.5$ values the green does not necessarily survive.

If we increase $q$ above a threshold value, a new phase emerges where all four strategies coexist. Starting from a random initial state, a representative time evolution is shown in panel~(b). Similarly to the previously described phase here green individuals almost die out first. But after a transient period the higher $q$ value offers them a chance to survive. Comparing to the previous panel, we can observe that the mentioned transient period is significantly shorter and it allows just a reduced oscillation of cyclic species before they reach their stationary portions. It simply means that the presence of an external predator can regulate the wild oscillations originated from a cyclic dominance. Our other observation is the unequal portions of cyclical species in the stationary state. As we already noted, here the presence of green species gives a slight benefit to blue individuals. This effect is not against the broadly valid and confirmed survival of the weakest effect because in our model the intensity of predations around the circle remained uniform for all species, therefore all predator-prey competitions will happen with the same likelihood. Instead, the presence of the green species results in fewer red individuals which lower the chance of a red--blue species competition. Therefore, the blue prey individuals do not die so often, while they can still benefit from the neighborhood of yellow fellows. 
Put differently, the survival of pestilent green species also breaks the delicate balance of original red-blue-yellow species. Therefore the competition followed by a reproduction process of May-Leonard model becomes less automatic as for smaller $q$ values. Consequently, the previously observed predation followed by a reproduction of the same predator happens not as often as for lower $q$ values, which gives a higher frequency of empty sites. In this way, not only $\rho_0$ increases, but also the border lines separating the spiral arms of cyclical species become less regular. We note that similar effect can be obtained by increasing mobility as it was shown in Ref.~\cite{2007-Reichenbach-N-488-1046}. In other words, the resulting mixed pattern does not support the above mentioned consecutive events, hence the increment of empty sites is another face of the fact that we loose the stable rotating spiral patterns.

If we strengthen the power of green species further by increasing $q$ then they may reach an even higher stationary portion, as shown in panel~(c). The argument we mentioned for the increment of blue species remains still valid, resulting that almost half of the available sites are occupied by them. A qualitatively new feature is the extinction of yellow species, as we already mentioned about Fig.~\ref{fig2}(c). We note that the decay of red species portion is not just useful for blue species, but is also harmful for yellow individuals because they simply loose their potential preys. More precisely, the direct green $\to$ red Lotka-Volterra-type invasion prevents them to produce empty space for a reproduction. This effect was already observed in panel~(b) of Fig.~\ref{fig3}, but it becomes dramatic for higher $q$, as shown in panel~(c).

Our last panel in Fig.~\ref{fig3} shows the extreme case when green species become too strong, or one might say too greedy. As a consequence, red and yellow species die out very fast and our pestilent predator is left out of food. They, however, decay very slowly, but permanently in time thanks to the low $d=1-q$ value. Therefore blue species, which is not vulnerable to the presence of green species, will gradually crowd out the latter resulting in a full-blue destination.

\begin{figure}[t]
	\centering
		\includegraphics[width= 7.9cm]{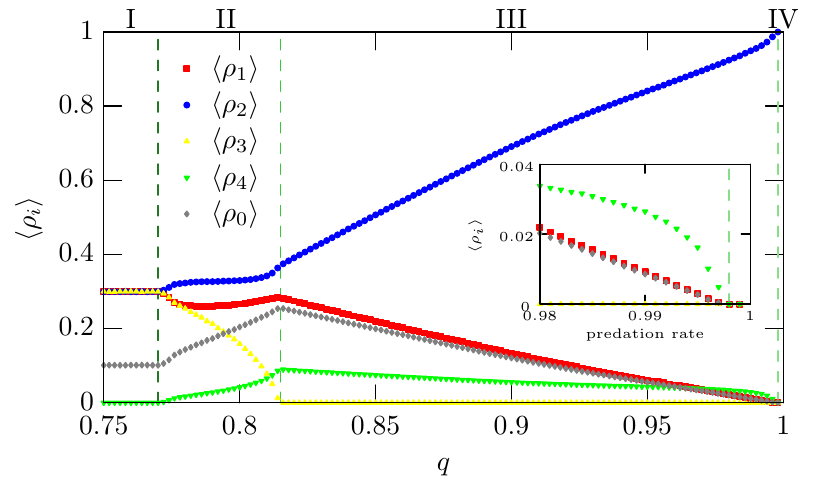}
	\caption{The average stationary density of species and empty sites as a function of the control parameter, which is the predation rate of the external green species. This plot reveals three consecutive phase transitions, marked by dashed vertical lines,  and four different phases in dependence of $q$. In Phase~I only cyclic species remain alive, while in Phase~II all competing species coexist. If we support green by larger $q$ then yellow individuals die out while the remaining three species still coexist in Phase~III. Last, for very high $q$ the system terminates to a uniform blue state, called as Phase~IV. The inset displays the vicinity of the last transition point where the external predator and its direct prey go extinct simultaneously.}
	\label{fig4}
\end{figure}

To quantify the borders of different phases in the whole $q$ interval we determined the stationary portions of all species and vacant sites for different $q$ values. 
To measure the stationary values properly we run the simulations for $15000$ MC steps where the values of the first $5000$ steps were discarded. Our results are summarized in Fig. \ref{fig4}. It shows clearly that we can detect three continuous phase transitions in dependence of the control parameter.

For low $q$ values of Phase~I only the cyclical species survive and we get back the traditional three-species model where the typical spatial patterns are illustrated in Fig.~\ref{fig2}{\rm(a)}. As we noted, the relatively small $q$ value cannot compensate the natural decay of green species. This situation changes above $q=0.770$ where all four species coexist. The mentioned transition point is marked by a vertical dashed line in Fig.~\ref{fig4}. The most striking feature of Phase~II is the fast decay of yellow species who gets a powerful competitor who shares the same red prey. More precisely, both yellow and green species beat red species, but the green species occupies the empty position immediately due to the applied Lotka-Volterra dynamical rule. Therefore yellow individuals remain empty handed while they are still predated by blue individuals. Fig.~\ref{fig4} shows very clearly that the average density of red species remains practically constant in this phase, therefore the real reason of the fall of yellow species is not the small abundance of red individuals, but the presence of green one who wins the indirect territorial battle. The fast raise of empty sites is also conspicuous in this phase which signals indirectly that the original rotating spiral pattern changes and the released empty sites after the fall of yellow individuals cannot be occupied effectively by blue players because the patterns become mixed.

By increasing the control parameter further, yellow species dies out at $q=0.815$ and we enter to Phase~III. Interestingly, the fall of one of the cyclical members does not involve the collapse of biodiversity, as one might expect. Instead, the remaining three species form a stable coexistence without having an explicit cycle among them. At first sight it seems to be a chain-like food web which should terminate biodiversity. But this argument is inaccurate, because, as we will point out later, there is still an effective circle in a way how microscopic states replace each other. More precisely, in the stationary state blue species is destroyed by red one, but blue individual can still reproduce because the annihilation of green species offers some empty place to them. A red species is replaced by a green one, but it can also reproduce in the empty sites left by green and the annihilation of blue when it is predated during a red $\to$ blue process. Lastly, green can build a balance between competing processes quantified by $q$ and $d$ values. To illustrate the stability of the described state, their coexistence remains stable even after $10^6$ evolutionary steps. 

The most striking feature of Phase~III is the spectacular rise of blue species as we increase $q$, hence the predation power of green individuals. This seems to be a paradoxical system reaction because by increasing $q$ we not just make the green $\to$ red direct invasion more effective, but simultaneously we lower the annihilation rate of green fellows. Still, as Fig.~\ref{fig4} illustrates very clearly, blue species benefits the most from this intervention. The simultaneous reverse fall of red curve demonstrates that the main benefit of blue is the fall of their natural predator which is red one.

As we increase $q$ even further the predation process between red and green individuals becomes so effective which eventually results in the extinction of both members of this pair, hence leaving blue species alone. This phase transition happens at $q=0.998$, as marked by a dashed vertical line. The inset of Fig.~\ref{fig4} shows this enlarged part in detail where the portion of red species becomes smaller than green species' share before extinction. It signals that green individuals, who can leave long due to very small $d$ value, are simply starving and they should also die in the lack of their prey individuals.

It is worth noting that in contrast to the case of Phase~II the portion of empty sites decays in Phase~III. In the latter phase only red and green individuals can create empty room. While the first does it via a predation process against blue, the latter makes empty site by its own annihilation. Notably, both processes become rare as we increase $q$, which explains the decay of the empty sites.

Summing up our observations, the clear victor of the presence of green species is blue one which is the most abundant for $q\geq 0.77$ and prevails the whole system as we approach the $q=1$ border line. One may argue that this system reaction is against the well-known and frequently confirmed survival of the weakest effect where those species become the most abundant which directly suffers from our external intervention into the cyclic dynamics. In our present case, however, there is a conceptual difference. Namely, we here maintain the uniformly strong predation among the cyclic members and change a portion of a single member directly via an external species. Nevertheless, the paradoxical behavior of cyclic systems reveals in another form because not the directly supported external predator will gain the most from this intervention, but the one who is the prey of attacked group member.

As we noted, one of the most spectacular observations is the stable coexistence of chain-like food web in Phase~III. In this case we cannot observe a direct cycle and the resulting pattern shown in Fig.~\ref{fig2}(c) confirms the lack of rotating spirals which are clearly present in panel~(a) obtained in Phase~I. Instead, the spatial distribution of spaces is rather homogeneous and the fluctuations are small as shown in third panel of Fig.~\ref{fig3}. These facts inspired us to apply a mean-field approximation. In this approach the density of empty sites and different species are marked by $\rho_i$ from $i=0, \dots 4$ respectively, where $\sum_{i} \,\rho_i=1$. The equation system which describes the time evolution of these variables are

\begin{eqnarray}
\nonumber
\dot{\rho}_1 &=& r \rho_0\rho_1 - q\rho_4\rho_1,\\ \nonumber
\dot{\rho}_2 &=& r \rho_0\rho_2 - p\rho_1\rho_2,\\ \nonumber
\dot{\rho}_4 &=& q \rho_1\rho_4 - (1-q)\rho_4, \nonumber
\end{eqnarray}
and
\begin{equation}
\dot{\rho}_0 = p \rho_1\rho_2 +(1-q)\rho_4-r (\rho_1+\rho_2)\rho_0 .
\nonumber
\end{equation}

\begin{figure}[h]
	\centering
		\includegraphics[width= 7.9cm]{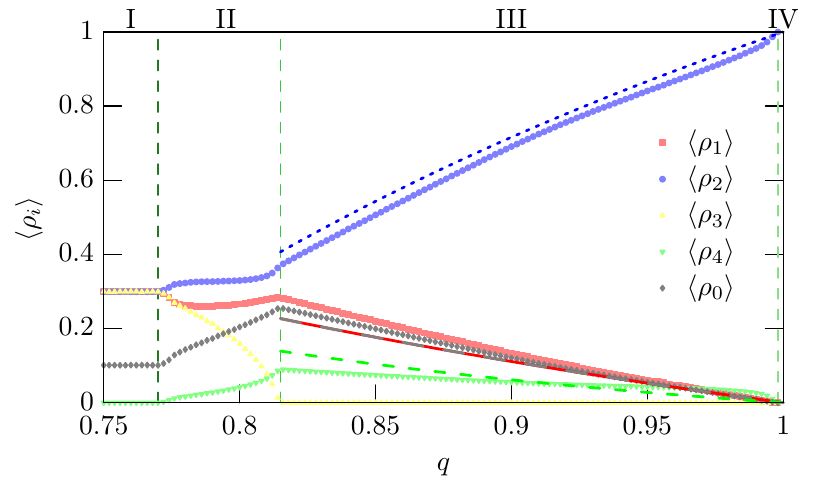}
	\caption{The comparison of simulation data (symbols) and predictions of mean-field approximation (appropriately colored dashed lines) in Phase~III when $r=p=0.5$. The two approaches show good qualitative agreement. Note that the predictions of $\rho_0$ and $\rho_1$ agree for these $r,p$ values hence the related lines collapse.}
	\label{Fig5}
\end{figure}

Please note that $\rho_3 = 0$ because we are in Phase~III.
The solutions for the stationary states are given by $\rho_1=(1-q)/q$, $\rho_2=1-[(1-q)/q][2+p/q]$, $\rho_4=(p/q)[(1-q)/q]$, and $\rho_0=(p/r)[(1-q)/q]$. For comparison we have plotted them by using appropriately colored dashed lines in Fig.~\ref{Fig5} in the case of $r=p=0.5$. This plot shows a nice qualitative agreement between the two approaches which indirectly supports our argument for the lack of spirals and the uniform spatial distribution, which makes such an approximation reasonable. 

The absence of direct cyclic dominance among green, red, and blue species in Phase~III and the related chain-like food-web make their coexistence a bit cryptic. But, as we already argued verbally, an effective dominance may still emerge among them, which can explain the stability of their association. To quantify this phenomenon, in the following we study the behavior of the Hamming distance density (HDD). Previous works of Bazeia {\it et al.} have pointed out that Hamming distance density characterizes heuristically the chaotic behavior in a cyclically dominant system and the amplitude of this quantity can be used as an order parameter to quantify chaos due to the fluctuations caused by non-transitive relation of competing members \cite{2017-Bazeia-SR-7-44900,2017-Bazeia-EPL-119-58003}. Therefore we measured HDD in the mentioned phase for three characteristic $q$ values which span the whole interval of Phase~III uniformly. These are $q=0.85, 0.9$, and $0.95$. 

In order to implement the numerical simulations, we prepared the initial state as follows: We distributed the red, blue, and green species and the empty sites randomly on the lattice, and applied $m=0.5$, $r=p=0.25$, $d=1-q$ parameter values for all mentioned $q$ values. In each case, we evolved the system for $5000$ MC steps and saved the resulting snapshot to be considered as the initial reference state. We then made a copy of it and modified the distribution by changing only the status of a single lattice site. We then evolved both the original and the slightly modified states by following exactly the same dynamical rules and monitor the corresponding Hamming distance values. For a reliable statistic we repeated this procedure $100$ times and averaged their values.

\begin{figure}{h}
	\centering
	\includegraphics[width= 7.9cm]{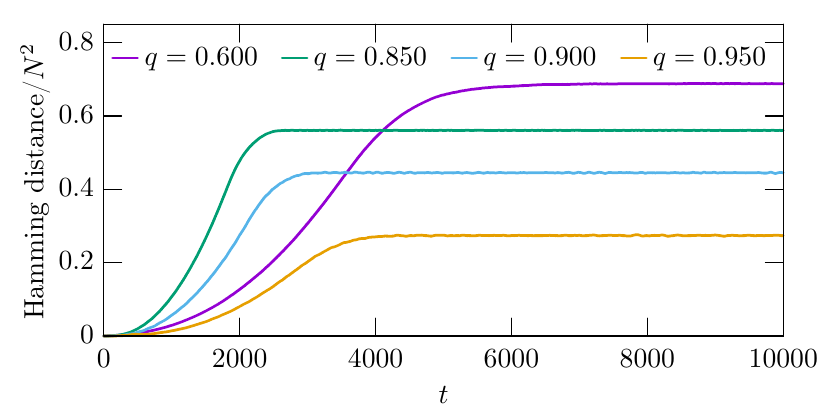}
	\caption{Time evolution of the Hamming distance density for $q$ values represent Phase~III as indicated in the legend. For comparison we also plotted this quantity for $q=0.6$ where the system evolves to the traditional three-species cyclically dominant model.}
	\label{Fig6}
\end{figure}

Our results are summarized in Fig.~\ref{Fig6} where, for comparison, we also plotted the case obtained at $q=0.6$. The latter value is from Phase~I where the system evolves into a clear cyclically dominant state. The comparison shows that HDD in Phase~III has very similar general time profile found earlier in cyclically dominant systems \cite{2017-Bazeia-SR-7-44900,2017-Bazeia-EPL-119-58003}. They suggest that there is still an effective cyclic transform among the microscopic states represented by different species. Indeed, the amplitude of HDD is the highest in the pure rock-paper-scissors model of Phase~I and its value decreases gradually as we approach Phase~IV. Evidently, the invasion flow among the surviving species is not as intensive as in Phase~I, especially for larger $q$ values, but this modest value can still explain why chain-like food web of green, red, and blue species can coexist. 

In the rest of this work we generalize our model to explore the robustness of our observations. More precisely, while we keep model parameters $m=0.5$, $r=p=0.25$ and $q=1-d$ unchanged, we extend the available neighborhood of the pestilent individuals. This extension was inspired by a recent work where the nearest, next-nearest and next-to-next-nearest neighbors were also included in a model study \cite{nosso}. This paper stressed that such an extension can increase the characteristic length of the system in a specific way, which could be vital for the resulting pattern formation. As we already noted in our present model, the spatial distribution of competing species could be decisive to reach the stable solution at certain parameter values, therefore the suggested extension could be useful tool to check the robustness of our previous findings.

In the modified model, when the neighborhood of a pestilent  individual is explored, we swap not only the four nearest-neighbors, but also add the eight next-nearest neighbor positions around the focal green fellow. It simply means that the latter player has now larger mobility and may hunt its red prey a little farther than before. The results of the modified model are similar to the ones displayed in Fig.~\ref{fig4}, with the previously observed three phase transitions remaining intact, but now with their positions shifted to slightly smaller values, giving by $q=0.744$, $0.800$, and $0.996$, respectively. 

\begin{figure}
	\centering
	\includegraphics[width= 7.9cm]{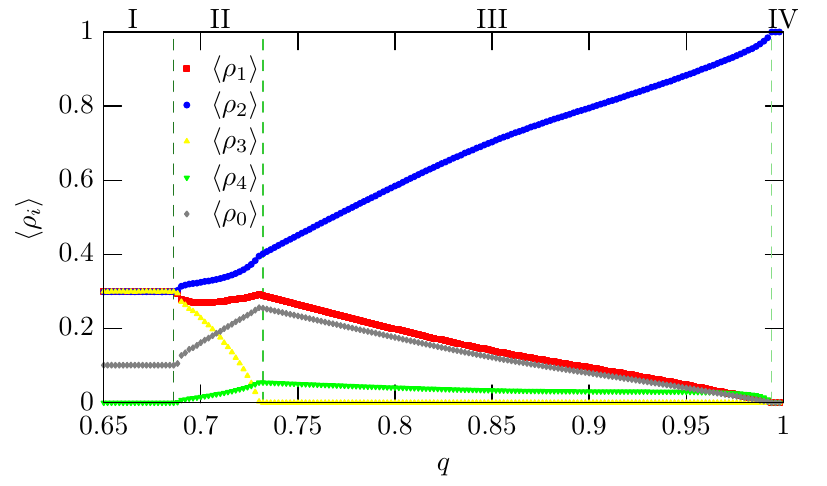}
	\caption{The average stationary density of the four species and empty sites in a modified model where the pestilent species have larger lethality. The positions of the phase transition points are marked by dashed vertical lines.}
	\label{fig7}
\end{figure}

We also considered an alternative modification of the original model where pestilent individuals may still interact with their nearest-neighbors but we increased their lethality. In particular, in contrast to the original model we allowed the focal green player to attack not only a single position among the neighbors, but the mentioned predator strikes two of the four positions simultaneously. More precisely, we introduced six combinations of the available pair of neighbors which are considered within a single attack. They are left-up (LU), left-right (LR), left-down (LD), up-right (UR), up-down (UD), and right-down (RD) pairs of nearest-neighbors. In this way the efficiency of the introduced Lotka-Volterra predation may be doubled despite the value of $q$. The resulting plot of the average densities of species is displayed in Fig.~\ref{fig7}, which highlights that the key features of the modified model remain intact. Interestingly, the positions of the phase transitions changed to lower values, which are now at $q=0.686$, $0.732$, and $0.994$. These observations suggest that the key feature which determines the model behavior is the topology of the complex food-web, and details of the microscopic spatial implementation or their strength have just second-order importance.

\section{Conclusion}

The main motivation of our present work was to check the stability of biodiversity build on cyclically dominant interactions of three competing species. In particular, we added a pestilent species which attacks only a single member of the mentioned triad. The proposed intervention is significantly different from previously introduced cases \cite{2017-Souza-Filho-PRE-95-062411,2018-Bazeia-EPL-124-68001,szolnoki_njp15} because the fundamental symmetry of the model is broken. Therefore we may expect that the presence of such kind of intruder may jeopardize not just a single species, but also the stability of the whole system. 

While using the standard parameters of May-Leonard models, such as mobility, reproduction, and competition, the control parameter of our new model is the $q$ predation strength of the pestilent species. Interestingly, we could detect three consecutive phase transitions as we increase the power of the mentioned predator. When this parameter is low then the external predator cannot compensate its natural annihilation rate and the system evolves to the traditional three-strategy model. But above a critical point all four species can survive and form a stable solution. By increasing $q$ further a member of the triad dies out, while the remaining three members can still coexist. Here we can list several interesting observations. First, the dying species is not the one that is attacked by the external predator, but its predator. In this way the balanced and closed food-web is broken and the interactions of remaining species change to a chain-like topology. This would terminate the biodiversity, but as we argued, an effective cyclical flow of microscopic states still exists which explains the coexistence. We directly measured the time evolution of Hamming distance density which confirmed our argument. Our third key observation is the paradoxical system behavior in this phase, where the heavier support of external predator will be harmful and results in a smaller average density of the mentioned competitor. This is again an interesting counter-intuitive consequence of a system where at least a latent cyclical flow is present. We note, however, that this is conceptually different from the well-known survival of the weakest effect or other previously reported phenomena \cite{tainaka_pla93,szolnoki_pre10b,szabo_jtb12}. As an extreme case of the above mentioned effect, if our pestilent species becomes too greedy then it may cause its end because it predates its prey too fast and after die out because of natural annihilation.

The robustness of our observations were also tested by extending the introduced model in different ways. In a version we extended the neighborhood size of pestilent individuals from which they collected their potential prey. In another version we doubled the lethality of the mentioned predator by attacking two neighbors simultaneously. Independently of the microscopic details the main findings remain intact and only the positions of the phase transitions change slightly.

This work may hopefully foster other investigations of current interest in several directions. In particular on the effects of the pestilent individuals in five or more cyclic species \cite{kang_pa13,vukov_pre13,park_c19b}, to see how the increment of the circular members  may change the emerging new phases. Another direction concerns the study of the system in an off-lattice environment, taking a continuum parameter to control the motion of the pestilent individuals, toward a more natural description of the real world.

Finally we note that our present model study pointed out that not only systems with direct cyclical dominances could provide counter-intuitive system behaviors. In certain cases it is enough to have just a second-order or latent circular flow of microscopic states to observe similarly surprising and undesigned effects. Therefore if we plan an encroachment into an ecosystem \cite{baker_jtb20}, an epidemic and spreading processes \cite{boerlijst_pcbi10,arruda_pr18} or change governing laws of societies where effective circles may also be present \cite{szolnoki_epl16,xia_cy_c20} then we should take special care of the subsequent behavior.

This work is supported in part by CAPES (Grant 88887.606927/2021-00), CNPq (Grants nos. 303469/2019-6 and 404913/2018-0), Para\'iba State Research Foundation (FAPESQ-PB, Grant no. 0015/2019), Funda\c c\~ao Arauc\'aria and INCT-FCx (CNPq/FAPESP).

\bibliographystyle{elsarticle-num-names}

\begin{thebibliography}{53}
\providecommand{\natexlab}[1]{#1}
\providecommand{\url}[1]{\texttt{#1}}
\providecommand{\urlprefix}{URL }
\expandafter\ifx\csname urlstyle\endcsname\relax
  \providecommand{\doi}[1]{doi:\discretionary{}{}{}#1}\else
  \providecommand{\doi}[1]{doi:\discretionary{}{}{}\begingroup
  \urlstyle{rm}\url{#1}\endgroup}\fi
\providecommand{\bibinfo}[2]{#2}

\bibitem[{Sinervo and Lively(1996)}]{1996-Sinervo-Nature-380-240}
\bibinfo{author}{B.~Sinervo}, \bibinfo{author}{C.~M. Lively},
  \bibinfo{title}{The rock-paper-scissors game and the evolution of alternative
  male strategies}, \bibinfo{journal}{Nature}
  \bibinfo{volume}{380}~(\bibinfo{number}{6571}) (\bibinfo{year}{1996})
  \bibinfo{pages}{240--243}.

\bibitem[{Kerr et~al.(2002)Kerr, Riley, Feldman, and
  Bohannan}]{2002-Kerr-N-418-171}
\bibinfo{author}{B.~Kerr}, \bibinfo{author}{M.~A. Riley},
  \bibinfo{author}{M.~W. Feldman}, \bibinfo{author}{B.~J.~M. Bohannan},
  \bibinfo{title}{Local dispersal promotes biodiversity in a real-life game of
  rock-paper-scissors}, \bibinfo{journal}{Nature}
  \bibinfo{volume}{418}~(\bibinfo{number}{6894}) (\bibinfo{year}{2002})
  \bibinfo{pages}{171--174}.

\bibitem[{Kirkup and Riley(2004)}]{2004-Kirkup-Nature-428-412}
\bibinfo{author}{B.~C. Kirkup}, \bibinfo{author}{M.~A. Riley},
  \bibinfo{title}{Antibiotic-mediated antagonism leads to a bacterial game of
  rock{\textendash}paper{\textendash}scissors in vivo},
  \bibinfo{journal}{Nature} \bibinfo{volume}{428}~(\bibinfo{number}{6981})
  (\bibinfo{year}{2004}) \bibinfo{pages}{412--414}.

\bibitem[{Reichenbach et~al.(2007)Reichenbach, Mobilia, and
  Frey}]{2007-Reichenbach-N-488-1046}
\bibinfo{author}{T.~Reichenbach}, \bibinfo{author}{M.~Mobilia},
  \bibinfo{author}{E.~Frey}, \bibinfo{title}{Mobility promotes and jeopardizes
  biodiversity in rock-paper-scissors games}, \bibinfo{journal}{Nature}
  \bibinfo{volume}{448}~(\bibinfo{number}{7157}) (\bibinfo{year}{2007})
  \bibinfo{pages}{1046--1049}.

\bibitem[{Song et~al.(2009)Song, Payne, Gray, and You}]{DB3}
\bibinfo{author}{H.~Song}, \bibinfo{author}{S.~Payne},
  \bibinfo{author}{M.~Gray}, \bibinfo{author}{L.~You},
  \bibinfo{title}{Spatiotemporal modulation of biodiversity in a synthetic
  chemical-mediated ecosystem}, \bibinfo{journal}{Nature Chemical Biology}
  \bibinfo{volume}{5}~(\bibinfo{number}{12}) (\bibinfo{year}{2009})
  \bibinfo{pages}{929--935}.

\bibitem[{Lozupone et~al.(2012)Lozupone, Stombaugh, Gordon, Jansson, and
  Knight}]{DB2}
\bibinfo{author}{C.~A. Lozupone}, \bibinfo{author}{J.~I. Stombaugh},
  \bibinfo{author}{J.~I. Gordon}, \bibinfo{author}{J.~K. Jansson},
  \bibinfo{author}{R.~Knight}, \bibinfo{title}{Diversity, stability and
  resilience of the human gut microbiota}, \bibinfo{journal}{Nature}
  \bibinfo{volume}{489}~(\bibinfo{number}{7415}) (\bibinfo{year}{2012})
  \bibinfo{pages}{220--230}.

\bibitem[{Levine et~al.(2017)Levine, Bascompte, Adler, and Allesina}]{DB1}
\bibinfo{author}{J.~M. Levine}, \bibinfo{author}{J.~Bascompte},
  \bibinfo{author}{P.~B. Adler}, \bibinfo{author}{S.~Allesina},
  \bibinfo{title}{Beyond pairwise mechanisms of species coexistence in complex
  communities}, \bibinfo{journal}{Nature}
  \bibinfo{volume}{546}~(\bibinfo{number}{7656}) (\bibinfo{year}{2017})
  \bibinfo{pages}{56--64}.

\bibitem[{Dobramysl et~al.(2018)Dobramysl, Mobilia, Pleimling, and
  Täuber}]{2018-Dobramysl-JPA-51-063001}
\bibinfo{author}{U.~Dobramysl}, \bibinfo{author}{M.~Mobilia},
  \bibinfo{author}{M.~Pleimling}, \bibinfo{author}{U.~C. Täuber},
  \bibinfo{title}{Stochastic population dynamics in spatially extended
  predator–prey systems}, \bibinfo{journal}{Journal of Physics A:
  Mathematical and Theoretical} \bibinfo{volume}{51}~(\bibinfo{number}{6})
  (\bibinfo{year}{2018}) \bibinfo{pages}{063001}.

\bibitem[{Szolnoki et~al.(2020)Szolnoki, de~Oliveira, and
  Bazeia}]{2020-Szolnoki-EPL-131-68001}
\bibinfo{author}{A.~Szolnoki}, \bibinfo{author}{B.~F. de~Oliveira},
  \bibinfo{author}{D.~Bazeia}, \bibinfo{title}{Pattern formations driven by
  cyclic interactions: A brief review of recent developments},
  \bibinfo{journal}{{EPL} (Europhysics Letters)}
  \bibinfo{volume}{131}~(\bibinfo{number}{6}) (\bibinfo{year}{2020})
  \bibinfo{pages}{68001}.

\bibitem[{Szab{\'{o}} and F{\'{a}}th(2007)}]{2007-Szabo-PR-446-97}
\bibinfo{author}{G.~Szab{\'{o}}}, \bibinfo{author}{G.~F{\'{a}}th},
  \bibinfo{title}{Evolutionary games on graphs}, \bibinfo{journal}{Physics
  Reports} \bibinfo{volume}{446}~(\bibinfo{number}{4-6}) (\bibinfo{year}{2007})
  \bibinfo{pages}{97--216}.

\bibitem[{Szolnoki et~al.(2014)Szolnoki, Mobilia, Jiang, Szczesny, Rucklidge,
  and Perc}]{2014-Szolnoki-JRSI-11-0735}
\bibinfo{author}{A.~Szolnoki}, \bibinfo{author}{M.~Mobilia},
  \bibinfo{author}{L.-L. Jiang}, \bibinfo{author}{B.~Szczesny},
  \bibinfo{author}{A.~M. Rucklidge}, \bibinfo{author}{M.~Perc},
  \bibinfo{title}{Cyclic dominance in evolutionary games: a review},
  \bibinfo{journal}{Journal of The Royal Society Interface}
  \bibinfo{volume}{11}~(\bibinfo{number}{100}) (\bibinfo{year}{2014})
  \bibinfo{pages}{20140735}.

\bibitem[{Mobilia et~al.(2016)Mobilia, Rucklidge, and Szczesny}]{mobilia_g16}
\bibinfo{author}{M.~Mobilia}, \bibinfo{author}{A.~M. Rucklidge},
  \bibinfo{author}{B.~Szczesny}, \bibinfo{title}{The Influence of Mobility Rate
  on Spiral Waves in Spatial Rock-Paper-Scissors Games},
  \bibinfo{journal}{Games} \bibinfo{volume}{7} (\bibinfo{year}{2016})
  \bibinfo{pages}{24}.

\bibitem[{Nagatani et~al.(2019)Nagatani, Ichinose, and
  i.~Tainaka}]{nagatani_jtb19}
\bibinfo{author}{T.~Nagatani}, \bibinfo{author}{G.~Ichinose},
  \bibinfo{author}{K.~i.~Tainaka}, \bibinfo{title}{Metapopulation dynamics in
  the rock-paper-scissors game with mutation: Effects of time-varying migration
  paths}, \bibinfo{journal}{J. Theor. Biol.} \bibinfo{volume}{462}
  (\bibinfo{year}{2019}) \bibinfo{pages}{425--431}.

\bibitem[{Szolnoki and Perc(2016{\natexlab{a}})}]{szolnoki_pre16}
\bibinfo{author}{A.~Szolnoki}, \bibinfo{author}{M.~Perc},
  \bibinfo{title}{Zealots tame oscillations in the spatial rock-paper-scissors
  game}, \bibinfo{journal}{Phys. Rev. E} \bibinfo{volume}{93}
  (\bibinfo{year}{2016}{\natexlab{a}}) \bibinfo{pages}{062307}.

\bibitem[{Brown et~al.(2019)Brown, Meyer-Ortmanns, and Pleimling}]{brown_pre19}
\bibinfo{author}{B.~L. Brown}, \bibinfo{author}{H.~Meyer-Ortmanns},
  \bibinfo{author}{M.~Pleimling}, \bibinfo{title}{Dynamically generated
  hierarchies in games of competition}, \bibinfo{journal}{Phys. Rev. E}
  \bibinfo{volume}{99} (\bibinfo{year}{2019}) \bibinfo{pages}{062116}.

\bibitem[{Park et~al.(2018)Park, Do, and Jang}]{park_c18c}
\bibinfo{author}{J.~Park}, \bibinfo{author}{Y.~Do}, \bibinfo{author}{B.~Jang},
  \bibinfo{title}{Multistability in the cyclic competition system},
  \bibinfo{journal}{Chaos} \bibinfo{volume}{28} (\bibinfo{year}{2018})
  \bibinfo{pages}{113110}.

\bibitem[{Garde et~al.(2020)Garde, Ewald, Kov{\'a}cs, and
  Schuster}]{garde_rsob20}
\bibinfo{author}{R.~Garde}, \bibinfo{author}{J.~Ewald},
  \bibinfo{author}{{\'A}.~T. Kov{\'a}cs}, \bibinfo{author}{S.~Schuster},
  \bibinfo{title}{Modelling population dynamics in a unicellular social
  organism community using a minimal model and evolutionary game theory},
  \bibinfo{journal}{Open Biol.} \bibinfo{volume}{10} (\bibinfo{year}{2020})
  \bibinfo{pages}{200206}.

\bibitem[{Palombi et~al.(2020)Palombi, Ferriani, and Toti}]{palombi_epjb20}
\bibinfo{author}{F.~Palombi}, \bibinfo{author}{S.~Ferriani},
  \bibinfo{author}{S.~Toti}, \bibinfo{title}{Coevolutionary dynamics of a
  variant of the cyclic \protect{Lotka-Volterra} model with three-agent
  interactions}, \bibinfo{journal}{Eur. Phys. J. B} \bibinfo{volume}{93}
  (\bibinfo{year}{2020}) \bibinfo{pages}{194}.

\bibitem[{Nagatani and Ichinose(2020)}]{nagatani_c20}
\bibinfo{author}{T.~Nagatani}, \bibinfo{author}{G.~Ichinose},
  \bibinfo{title}{Diffusively-Coupled Rock-Paper-Scissors Game with Mutation in
  Scale-Free Hierarchical Networks}, \bibinfo{journal}{Complexity}
  \bibinfo{volume}{2020} (\bibinfo{year}{2020}) \bibinfo{pages}{6976328}.

\bibitem[{Szolnoki and Perc(2016{\natexlab{b}})}]{szolnoki_srep16b}
\bibinfo{author}{A.~Szolnoki}, \bibinfo{author}{M.~Perc},
  \bibinfo{title}{Biodiversity in models of cyclic dominance is preserved by
  heterogeneity in site-specific invasion rates}, \bibinfo{journal}{Sci. Rep.}
  \bibinfo{volume}{6} (\bibinfo{year}{2016}{\natexlab{b}})
  \bibinfo{pages}{38608}.

\bibitem[{Frean and Abraham(2001)}]{2001-Frean-PRSLB-268-1323}
\bibinfo{author}{M.~Frean}, \bibinfo{author}{E.~R. Abraham},
  \bibinfo{title}{Rock-scissors-paper and the survival of the weakest},
  \bibinfo{journal}{Proc. R. Soc. Lond. B}
  \bibinfo{volume}{268}~(\bibinfo{number}{1474}) (\bibinfo{year}{2001})
  \bibinfo{pages}{1323--1327}.

\bibitem[{Tainaka(1993)}]{tainaka_pla93}
\bibinfo{author}{K.~Tainaka}, \bibinfo{title}{Paradoxial effect in a
  three-candidate voter model}, \bibinfo{journal}{Phys. Lett. A}
  \bibinfo{volume}{176} (\bibinfo{year}{1993}) \bibinfo{pages}{303--306}.

\bibitem[{Berr et~al.(2009)Berr, Reichenbach, Schottenloher, and
  Frey}]{Berr-PRL-102-048102}
\bibinfo{author}{M.~Berr}, \bibinfo{author}{T.~Reichenbach},
  \bibinfo{author}{M.~Schottenloher}, \bibinfo{author}{E.~Frey},
  \bibinfo{title}{Zero-One Survival Behavior of Cyclically Competing Species},
  \bibinfo{journal}{Phys. Rev. Lett.} \bibinfo{volume}{102}
  (\bibinfo{year}{2009}) \bibinfo{pages}{048102}.

\bibitem[{Blahota et~al.(2020)Blahota, Blahota, and Szolnoki}]{blahota_epl20}
\bibinfo{author}{M.~Blahota}, \bibinfo{author}{I.~Blahota},
  \bibinfo{author}{A.~Szolnoki}, \bibinfo{title}{Equal partners do better in
  defensive alliances}, \bibinfo{journal}{EPL} \bibinfo{volume}{131}
  (\bibinfo{year}{2020}) \bibinfo{pages}{58002}.

\bibitem[{Avelino et~al.(2019)Avelino, de~Oliveira, and
  Trintin}]{2019-Avelino-PRE-100-042209}
\bibinfo{author}{P.~P. Avelino}, \bibinfo{author}{B.~F. de~Oliveira},
  \bibinfo{author}{R.~S. Trintin}, \bibinfo{title}{Predominance of the weakest
  species in Lotka-Volterra and May-Leonard formulations of the
  rock-paper-scissors model}, \bibinfo{journal}{Phys. Rev. E}
  \bibinfo{volume}{100} (\bibinfo{year}{2019}) \bibinfo{pages}{042209}.

\bibitem[{Szolnoki and Chen(2020)}]{szolnoki_csf20b}
\bibinfo{author}{A.~Szolnoki}, \bibinfo{author}{X.~Chen},
  \bibinfo{title}{Strategy dependent learning activity in cyclic dominant
  systems}, \bibinfo{journal}{Chaos Soliton. Fract.} \bibinfo{volume}{138}
  (\bibinfo{year}{2020}) \bibinfo{pages}{109935}.

\bibitem[{Nagatani(2019)}]{nagatani_pa19b}
\bibinfo{author}{T.~Nagatani}, \bibinfo{title}{Diffusively coupled
  Lotka-Volterra system stabilized by heterogeneous graphs},
  \bibinfo{journal}{Physica A} \bibinfo{volume}{525} (\bibinfo{year}{2019})
  \bibinfo{pages}{1114--1123}.

\bibitem[{Avelino et~al.(2020)Avelino, de~Oliveira, and
  Trintin}]{2020-Avelino-PRE-101-062312}
\bibinfo{author}{P.~P. Avelino}, \bibinfo{author}{B.~F. de~Oliveira},
  \bibinfo{author}{R.~S. Trintin}, \bibinfo{title}{Performance of weak species
  in the simplest generalization of the rock-paper-scissors model to four
  species}, \bibinfo{journal}{Phys. Rev. E} \bibinfo{volume}{101}
  (\bibinfo{year}{2020}) \bibinfo{pages}{062312}.

\bibitem[{Szolnoki and Chen(2017)}]{szolnoki_pre17}
\bibinfo{author}{A.~Szolnoki}, \bibinfo{author}{X.~Chen},
  \bibinfo{title}{Alliance formation with exclusion in the spatial public goods
  game}, \bibinfo{journal}{Phys. Rev. E} \bibinfo{volume}{95}
  (\bibinfo{year}{2017}) \bibinfo{pages}{052316}.

\bibitem[{Park(2018)}]{park_c18}
\bibinfo{author}{J.~Park}, \bibinfo{title}{Biodiversity in the cyclic
  competition system of three species according to the emergence of mutant
  species}, \bibinfo{journal}{Chaos} \bibinfo{volume}{28}
  (\bibinfo{year}{2018}) \bibinfo{pages}{053111}.

\bibitem[{Bazeia et~al.(2020)Bazeia, de~Oliveira, Silva, and
  Szolnoki}]{2020-Bazeia-CSF-141-110356}
\bibinfo{author}{D.~Bazeia}, \bibinfo{author}{B.~de~Oliveira},
  \bibinfo{author}{J.~Silva}, \bibinfo{author}{A.~Szolnoki},
  \bibinfo{title}{Breaking unidirectional invasions jeopardizes biodiversity in
  spatial May-Leonard systems}, \bibinfo{journal}{Chaos, Solitons {\&}
  Fractals} \bibinfo{volume}{141} (\bibinfo{year}{2020})
  \bibinfo{pages}{110356}.

\bibitem[{Wang et~al.(2014)Wang, Szolnoki, and Perc}]{wang_z_pre14b}
\bibinfo{author}{Z.~Wang}, \bibinfo{author}{A.~Szolnoki},
  \bibinfo{author}{M.~Perc}, \bibinfo{title}{Different perceptions of social
  dilemmas: Evolutionary multigames in structured populations},
  \bibinfo{journal}{Phys. Rev. E} \bibinfo{volume}{90} (\bibinfo{year}{2014})
  \bibinfo{pages}{032813}.

\bibitem[{Liao et~al.(2020)Liao, Miano, Nguyen, Chao, and Hasty}]{Liao2020}
\bibinfo{author}{M.~J. Liao}, \bibinfo{author}{A.~Miano},
  \bibinfo{author}{C.~B. Nguyen}, \bibinfo{author}{L.~Chao},
  \bibinfo{author}{J.~Hasty}, \bibinfo{title}{Survival of the weakest in
  non-transitive asymmetric interactions among strains of E. coli},
  \bibinfo{journal}{Nature Communications}
  \bibinfo{volume}{11}~(\bibinfo{number}{1}).

\bibitem[{Bazeia et~al.(2017)Bazeia, Pereira, Brito, de~Oliveira, and
  Ramos}]{2017-Bazeia-SR-7-44900}
\bibinfo{author}{D.~Bazeia}, \bibinfo{author}{M.~B. P.~N. Pereira},
  \bibinfo{author}{A.~V. Brito}, \bibinfo{author}{B.~de~Oliveira},
  \bibinfo{author}{J.~G. G.~S. Ramos}, \bibinfo{title}{A novel procedure for
  the identification of chaos in complex biological systems},
  \bibinfo{journal}{Scientific Reports} \bibinfo{volume}{7}
  (\bibinfo{year}{2017}) \bibinfo{pages}{44900}.

\bibitem[{Hamming(1950)}]{Hamming}
\bibinfo{author}{R.~W. Hamming}, \bibinfo{title}{Error detecting and error
  correcting codes}, \bibinfo{journal}{The Bell system technical journal}
  \bibinfo{volume}{29}~(\bibinfo{number}{2}) (\bibinfo{year}{1950})
  \bibinfo{pages}{147--160}.

\bibitem[{{Bazeia, D.} et~al.(2017){Bazeia, D.}, {Menezes, J.}, {de Oliveira,
  B. F.}, and {Ramos, J. G. G. S.}}]{2017-Bazeia-EPL-119-58003}
\bibinfo{author}{{Bazeia, D.}}, \bibinfo{author}{{Menezes, J.}},
  \bibinfo{author}{{de Oliveira, B. F.}}, \bibinfo{author}{{Ramos, J. G. G.
  S.}}, \bibinfo{title}{Hamming distance and mobility behavior in generalized
  rock-paper-scissors models} .

\bibitem[{Souza-Filho et~al.(2017)Souza-Filho, Bazeia, and
  Ramos}]{2017-Souza-Filho-PRE-95-062411}
\bibinfo{author}{C.~A. Souza-Filho}, \bibinfo{author}{D.~Bazeia},
  \bibinfo{author}{J.~G. G.~S. Ramos}, \bibinfo{title}{Apex predator and the
  cyclic competition in a rock-paper-scissors game of three species},
  \bibinfo{journal}{Phys. Rev. E} \bibinfo{volume}{95} (\bibinfo{year}{2017})
  \bibinfo{pages}{062411}.

\bibitem[{Bazeia et~al.(2018)Bazeia, de~Oliveira, and
  Szolnoki}]{2018-Bazeia-EPL-124-68001}
\bibinfo{author}{D.~Bazeia}, \bibinfo{author}{B.~F. de~Oliveira},
  \bibinfo{author}{A.~Szolnoki}, \bibinfo{title}{Phase transitions in
  dependence of apex predator decaying ratio in a cyclic dominant system},
  \bibinfo{journal}{{EPL} (Europhysics Letters)}
  \bibinfo{volume}{124}~(\bibinfo{number}{6}) (\bibinfo{year}{2018})
  \bibinfo{pages}{68001}.

\bibitem[{Durney et~al.(2011)Durney, Case, Pleimling, and
  Zia}]{2011-Durney-PRE-83-051108}
\bibinfo{author}{C.~H. Durney}, \bibinfo{author}{S.~O. Case},
  \bibinfo{author}{M.~Pleimling}, \bibinfo{author}{R.~K.~P. Zia},
  \bibinfo{title}{Saddles, arrows, and spirals: Deterministic trajectories in
  cyclic competition of four species}, \bibinfo{journal}{Phys. Rev. E}
  \bibinfo{volume}{83} (\bibinfo{year}{2011}) \bibinfo{pages}{051108}.

\bibitem[{Knebel et~al.(2013)Knebel, Kr\"{u}ger, Weber, and
  Frey}]{2013-Knebel-PRL-110-168106}
\bibinfo{author}{J.~Knebel}, \bibinfo{author}{T.~Kr\"{u}ger},
  \bibinfo{author}{M.~F. Weber}, \bibinfo{author}{E.~Frey},
  \bibinfo{title}{Coexistence and Survival in Conservative Lotka-Volterra
  Networks}, \bibinfo{journal}{Phys. Rev. Lett.}
  \bibinfo{volume}{110}~(\bibinfo{number}{16}) (\bibinfo{year}{2013})
  \bibinfo{pages}{168106}.

\bibitem[{Brown and Pleimling(2017)}]{2017-Brown-PRE-96-012147}
\bibinfo{author}{B.~L. Brown}, \bibinfo{author}{M.~Pleimling},
  \bibinfo{title}{Coarsening with nontrivial in-domain dynamics: Correlations
  and interface fluctuations}, \bibinfo{journal}{Phys. Rev. E}
  \bibinfo{volume}{96} (\bibinfo{year}{2017}) \bibinfo{pages}{012147}.

\bibitem[{Frey(2010)}]{frey_pa10}
\bibinfo{author}{E. Frey},   \bibinfo{title}{Evolutionary game theory: Theoretical concepts and applications to
microbial communities}, \bibinfo{journal}{Physica A}
  \bibinfo{volume}{389} (\bibinfo{year}{2010}) \bibinfo{pages}{4265--4298}.

\bibitem[{Bazeia et~al.(2021)Bazeia, Bongestab, and \protect{de
  Oliveira}}]{nosso}
\bibinfo{author}{D.~Bazeia}, \bibinfo{author}{M.~Bongestab},
  \bibinfo{author}{B.~F. \protect{de Oliveira}}, \bibinfo{title}{Influence of
  the neighborhood on cyclic models of biodiversity},
  \bibinfo{journal}{preprint arXiv:2103.05040} .

\bibitem[{Szolnoki and Perc(2015)}]{szolnoki_njp15}
\bibinfo{author}{A.~Szolnoki}, \bibinfo{author}{M.~Perc},
  \bibinfo{title}{Vortices determine the dynamics of biodiversity in cyclical
  interactions with protection spillovers}, \bibinfo{journal}{New J. Phys.}
  \bibinfo{volume}{17} (\bibinfo{year}{2015}) \bibinfo{pages}{113033}.

\bibitem[{Szolnoki et~al.(2010)Szolnoki, Wang, Wang, and Zhu}]{szolnoki_pre10b}
\bibinfo{author}{A.~Szolnoki}, \bibinfo{author}{Z.~Wang},
  \bibinfo{author}{J.~Wang}, \bibinfo{author}{X.~Zhu},
  \bibinfo{title}{Dynamically generated cyclic dominance in spatial prisoner's
  dilemma games}, \bibinfo{journal}{Phys. Rev. E} \bibinfo{volume}{82}
  (\bibinfo{year}{2010}) \bibinfo{pages}{036110}.

\bibitem[{Szab{\'o} and Szolnoki(2012)}]{szabo_jtb12}
\bibinfo{author}{G.~Szab{\'o}}, \bibinfo{author}{A.~Szolnoki},
  \bibinfo{title}{Selfishness, fraternity, and other-regarding preference in
  spatial evolutionary games}, \bibinfo{journal}{J. Theor. Biol.}
  \bibinfo{volume}{299} (\bibinfo{year}{2012}) \bibinfo{pages}{81--87}.

\bibitem[{Kang et~al.(2013)Kang, Pan, Wang, and He}]{kang_pa13}
\bibinfo{author}{Y.~Kang}, \bibinfo{author}{Q.~Pan}, \bibinfo{author}{X.~Wang},
  \bibinfo{author}{M.~He}, \bibinfo{title}{A golden point rule in
  rock--paper--scissors--lizard--spock game}, \bibinfo{journal}{Physica A}
  \bibinfo{volume}{392} (\bibinfo{year}{2013}) \bibinfo{pages}{2652--2659}.

\bibitem[{Vukov et~al.(2013)Vukov, Szolnoki, and Szab{\'o}}]{vukov_pre13}
\bibinfo{author}{J.~Vukov}, \bibinfo{author}{A.~Szolnoki},
  \bibinfo{author}{G.~Szab{\'o}}, \bibinfo{title}{Diverging fluctuations in a
  spatial five-species cyclic dominance game}, \bibinfo{journal}{Phys. Rev. E}
  \bibinfo{volume}{88} (\bibinfo{year}{2013}) \bibinfo{pages}{022123}.

\bibitem[{Park and Jang(2019)}]{park_c19b}
\bibinfo{author}{J.~Park}, \bibinfo{author}{B.~Jang}, \bibinfo{title}{Robust
  coexistence with alternative competition strategy in the spatial cyclic game
  of five species}, \bibinfo{journal}{Chaos} \bibinfo{volume}{29}
  (\bibinfo{year}{2019}) \bibinfo{pages}{051105}.

\bibitem[{Baker and Pleimling(2020)}]{baker_jtb20}
\bibinfo{author}{R.~Baker}, \bibinfo{author}{M.~Pleimling}, \bibinfo{title}{The
  effect of habitats and fitness on species coexistence in systems with cyclic
  dominance}, \bibinfo{journal}{J. Theor. Biol.} \bibinfo{volume}{486}
  (\bibinfo{year}{2020}) \bibinfo{pages}{110084}.

\bibitem[{Boerlijst and van Ballegooijen(2010)}]{boerlijst_pcbi10}
\bibinfo{author}{M.~C. Boerlijst}, \bibinfo{author}{W.~M. van Ballegooijen},
  \bibinfo{title}{Spatial Pattern Switching Enables Cyclic Evolution in Spatial
  Epidemics}, \bibinfo{journal}{PLoS Comput. Biol.} \bibinfo{volume}{6}
  (\bibinfo{year}{2010}) \bibinfo{pages}{e1001030}.

\bibitem[{de~Arruda et~al.(2018)de~Arruda, Rodrigues, and Moreno}]{arruda_pr18}
\bibinfo{author}{G.~F. de~Arruda}, \bibinfo{author}{F.~A. Rodrigues},
  \bibinfo{author}{Y.~Moreno}, \bibinfo{title}{Fundamentals of spreading
  processes in single and multilayer complex networks}, \bibinfo{journal}{Phys.
  Rep.} \bibinfo{volume}{756} (\bibinfo{year}{2018}) \bibinfo{pages}{1--59}.

\bibitem[{Szolnoki and Perc(2016{\natexlab{c}})}]{szolnoki_epl16}
\bibinfo{author}{A.~Szolnoki}, \bibinfo{author}{M.~Perc},
  \bibinfo{title}{Collective influence in evolutionary social dilemmas},
  \bibinfo{journal}{EPL} \bibinfo{volume}{113}
  (\bibinfo{year}{2016}{\natexlab{c}}) \bibinfo{pages}{58004}.

\bibitem[{Xia et~al.(2020)Xia, Gracia-L{\'a}zaro, and Moreno}]{xia_cy_c20}
\bibinfo{author}{C.~Xia}, \bibinfo{author}{C.~Gracia-L{\'a}zaro},
  \bibinfo{author}{Y.~Moreno}, \bibinfo{title}{Effect of memory, intolerance,
  and second-order reputation on cooperation}, \bibinfo{journal}{Chaos}
  \bibinfo{volume}{30} (\bibinfo{year}{2020}) \bibinfo{pages}{063122}.

\end{thebibliography}

\end{document}